\documentclass[conference]{IEEEtran}
\usepackage{url}
\let\labelindent\relax
\usepackage{enumitem}
\setitemize[0]{leftmargin=*,partopsep=1pt,topsep=1pt}
\usepackage{verbatim}
\ifCLASSINFOpdf
   \usepackage[pdftex]{graphicx}
    \usepackage{subfig}
\else
\fi
\begin{document}
%
\title{MCUIUC -- A New Framework for Metagenomic Read Compression}

\author{\IEEEauthorblockN{Jonathan G. Ligo$^{*}$, Minji Kim$^{*}$, Amin Emad, Olgica Milenkovic and Venugopal V. Veeravalli\\
$^{*}$The first two authors contributed equally to this paper.}
\IEEEauthorblockA{Department of Electrical and Computer Engineering, University of Illinois, Urbana-Champaign\\
Email: \{ligo2,mkim158,emad2,milenkov,vvv\} [at] illinois.edu}}

\maketitle

\begin{abstract}
Metagenomics is an emerging field of molecular biology concerned with analyzing the genomes of environmental samples comprising many different diverse organisms.
Given the nature of metagenomic data, one usually has to sequence the genomic material of all organisms in a batch, leading to a mix of reads coming from different DNA
sequences. In deep high-throughput sequencing experiments, the volume of the raw reads is extremely high, frequently exceeding 600 Gb. With an ever increasing demand for storing 
such reads for future studies, the issue of efficient metagenomic compression becomes of paramount importance. We present
the first known approach to metagenome read compression, termed MCUIUC (Metagenomic Compression at UIUC). The gist of the proposed algorithm is to perform classification of reads based on unique organism identifiers, followed by reference-based alignment of reads for individually identified organisms, and metagenomic assembly of unclassified reads. Once assembly and classification are completed, lossless reference based compression is performed via positional encoding. We evaluate the performance of the algorithm on moderate sized synthetic metagenomic samples involving 15 randomly selected organisms and describe future directions for improving the proposed compression method.
\end{abstract}

\IEEEpeerreviewmaketitle

\section{Introduction} \label{sec:intro}
Metagenomics is an area of microbial genomics devoted to the unified study of complex cultures found in the human body and the ecosystem. The main difference between classical genomics and metagenomics is the fact that in the latter, no attempt is made to separate the organisms in a sample. The reasons for using such a holistic approach are either of purely technical nature -- in cases when single sample isolation is technically implausible to accomplish -- or they may be governed by the desire to explore the interactions, dynamical co-regulation and joint evolution of microbes in a given environment~\cite{R2004}.

Unlike the field of genomics, metagenomics is still in its infancy, despite recent intense research efforts in the areas of metagenomic sampling, sequencing and assembly~\cite{R2004,Q2010,Pengetal2012}. The main challenges in the field are a still fairly large cost of high-coverage metagenomic sequencing, although cost constraints are likely to diminish in the near future; and the extremely large file sizes produced by High-Throughput Sequencers such as Illumina, when applied to metagenomic data. As an illustration, the size of the human gut microbial population is estimated to be several hundreds, while average bacterial genome lengths
are between $100$ Kbp (Kilobasepairs) and $12$ Mbp (Megabasepairs). With a coverage of at least $100$ reads per base pair, one arrives at a rough estimate of $100$ Gbp (Gigabasepairs) of genomic data. Analyzing, storing and transmitting such volumes of data is a formidable task, and most efforts on metagenome assembly require powerful and parallelized computer clusters to perform even the most basic operations.

In parallel, an extensive effort is underway to develop efficient means for lossless and lossy compression of whole genomes~\cite{S2008}, as well as for compression of reads sampled from a single genome~\cite{M2011}. Specialized methods for compressing protein-coding regions as well as regions with frequent repeats were reported in~\cite{A2002,CMV_05,NM1999,X2002}, using methods as diverse as modified Lempel-Ziv encoding, the Burrows-Wheeler transform and wavelet-based decomposition. Compression of reads is mainly achieved by using a reference genome, with an initial study based on Golomb codes reported in~\cite{M2011} and extensions thereof reported in~\cite{T13}.

We propose the first algorithmic solution to the problem of \emph{de novo} metagenomic compression that mitigates the need for computationally demanding \emph{full} metagenomic assembly by using a read classification technique and subsequent reference-based single genome compression. The gist of the classification method is to use a microbe identification tool -- \emph{Metaphyler}~\cite{Liuetal2010} -- that can accurately identify the order of a high percentage of organisms from a metagenomic mix. By aligning the reads to \emph{all genomes} of organisms within the identified orders via \emph{Bowtie2}~\cite{LS2012}, one can provide a rough classification map for reference-based compression. For moderate size samples, unaligned reads usually amount to less than $20-25\%$ of the reads, and can be assembled through existing metagenome assembly software such as IDBA-UD~\cite{Pengetal2012}. Starting positions of aligned reads and their differences from the reference genome are encoded using techniques outlined in~\cite{M2011,cram}.

The paper is organized as follows. In Section~\ref{sec:algorithm}, we describe the first algorithmic solution for metagenomic read compression. In Section~\ref{sec:example}, we demonstrate the performance of the method on synthetic Illumina sequencer data, using a randomly selected set of $15$ bacterial organisms. Concluding remarks are given in Section~\ref{sec:conclusion}.

\section{Algorithmic Solution for Metagenomic Compression} \label{sec:algorithm}

\begin{figure}[!t]
\centering
\includegraphics[width=3.0in]{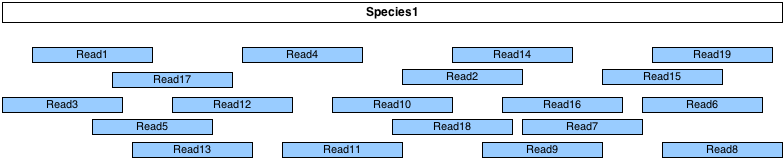}
\caption{Single genome alignment. Blue reads are aligned to their correct positions.}
\label{fig:single-genome}
\end{figure}

\begin{figure}[!t]
\centering
\includegraphics[width=3.0in]{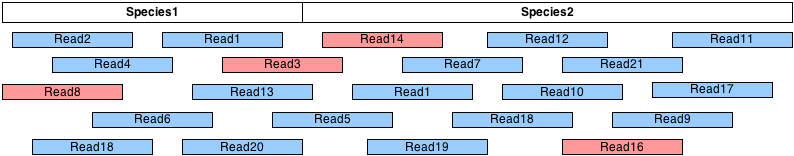}
\caption{Two genomes alignment. Blue reads are aligned to their correct positions in the correct genome. Red reads are aligned to the genome of the wrong species.}
\label{fig:two-genomes}
\vspace{-20pt}
\end{figure}

Throughout the paper, we refer to an ordered sequence of symbols from the alphabet $\{{A,T,G,C\}}$ as a \emph{genome} or \emph{genomic sequence}. Elements of the sequence are indexed from $1$ to $L$, the length of the sequence. 
A \emph{read} is a \emph{substring} of a genome, generated by some sequencing system. The starting and terminal location of a read correspond to the index of the first and last element of the read in the underlying genome. The support of the read is the set of indices of the genome between the starting and terminal location of a read. 
The \emph{coverage} of an element in a given genome equals the number of reads that contain the index of the element in their support. \emph{Assembly} refers to the process of overlapping reads -- suffix to prefix -- in 
order to reconstruct the original sequence the reads came from. \emph{Alignment} refers to mapping reads onto a given genome. An example of a genome and a set of reads aligned to the sequence are depicted in Figure~\ref{fig:single-genome}. A similar alignment involving
two genomes is depicted in Figure~\ref{fig:two-genomes}, also depicting several misaligned reads between species.

Given that metagenomic data carries valuable information about the population dynamics of microbes in a given environment, even small aberrations in the composition of the sample have to
be observed and recorded for in-depth study of the underlying phenomena. Also, given the cost of acquiring samples to sequence and laboratory work necessary for sequencing, it is also valuable to preserve and share the sequencing experiment in its entirety. Hence, the goal of any compression method should be to preserve the reads intact, i.e., to perform lossless read compression.
The main issues associated with individual lossless read compression is that the reads -- as generated by Illumina, Roche454, IonTorrent -- are usually of length not exceeding $L=400-500$ bps, and that the number of reads is extremely large. The former property of reads does not allow any known compression algorithm to take full advantage of inherent patterns or data structures, since such algorithms
offer good performance only for sufficiently long sequence lengths~\cite{CT}. On the other hand, the existence of a huge volume of reads creates the problem of efficiently using their overlaps.
Such overlaps arise due to the requirement of large genome coverage, given that large coverage is needed for accurate sequence reconstruction\footnote{What may be deemed sufficiently 
large coverage depends on many factors, including the error rate of the sequencer, the structural properties of the sequenced genomes and other properties. For example, some regions of a genome that contain many repeats or approximate repeats cannot be accurately sequenced no matter how large the coverage~\cite{S98}.}.

One simple method to mitigate the aforementioned problems is to perform \emph{reference based compression}~\cite{M2011}. Reference based compression is based on the assumption that one knows some rough form of the genome the reads were generated from, without knowing the exact identity of the organism. Such an approach cannot be used for \emph{de novo} compression,
where the origin of the reads is unknown. Such is the case for most metagenomic datasets currently available.

\begin{figure*}[t]
\begin{centering}
\centerline{\includegraphics[width=.72\textwidth]{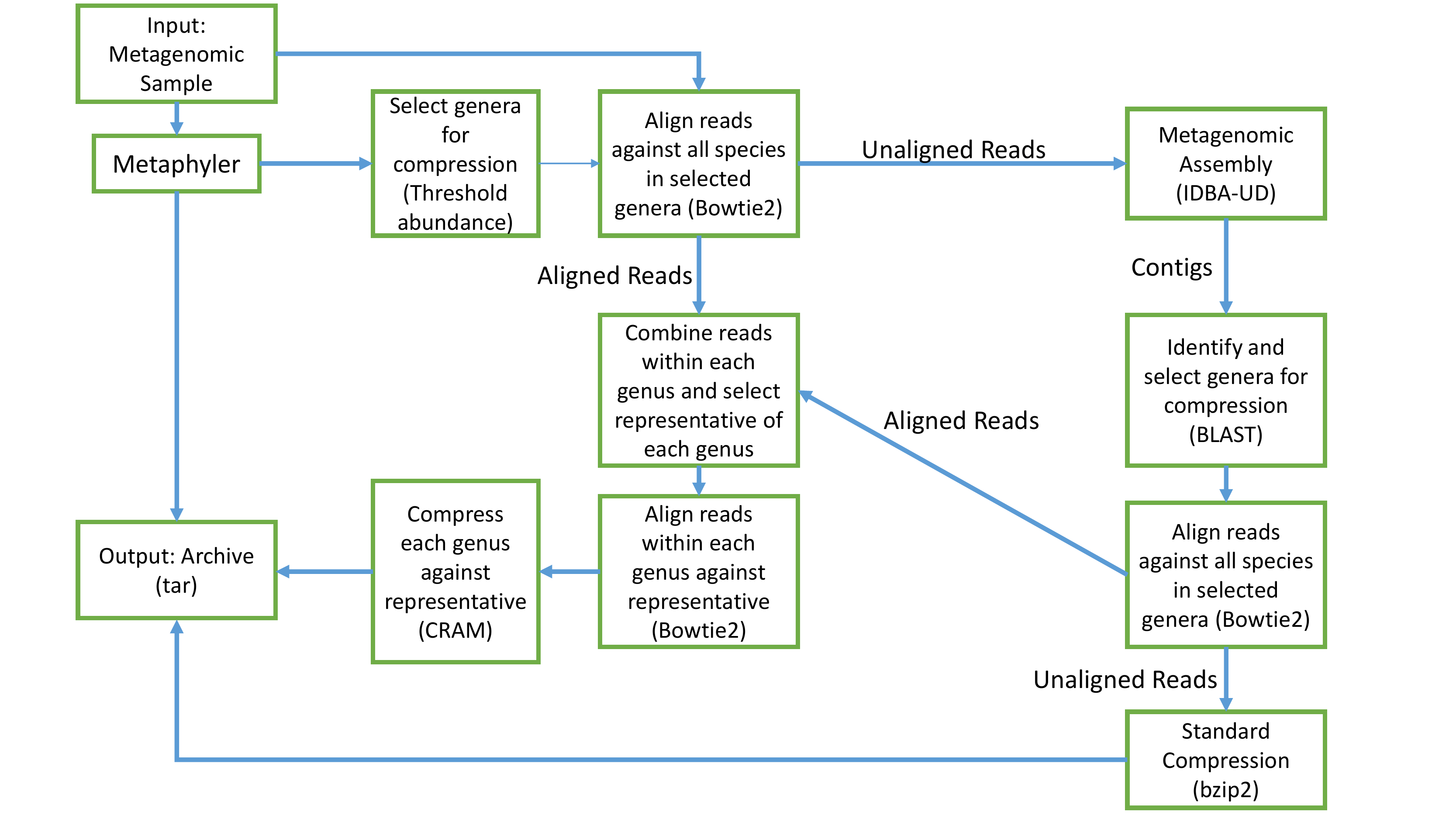}}
\caption{Block Diagram of the MCUIUC Algorithm for Lossless Metagenomic Compression.}\label{fig:MCUIUC}
\end{centering}
\vspace{-15pt} 
\end{figure*}

One obvious solution would be to first apply a metagenomic assembly process to create an adequate number of reference genomes, and subsequently compress all the reads using the ``closest'' genome as a reference. This is infeasible in the long term due to the reduction of costs in sequencing technology, increases in sequencing depth and a desire to sequence more complex metagenomic samples. A current Next Generation sequencer such as the Illumina HiSeq 2500 system can produce up to 4.5 Gb/hr~\cite{hiseq}, a rate which will be quickly exceeded by future sequencers. Thus, a laboratory running multiple sequencers must archive and distribute petabytes of data per year. Since the read lengths are short, techniques such as de Brujin graph assemblers~\cite{Pengetal2012} are needed for assembly at the cost of high memory and CPU time for large data sets, requiring prohibitively expensive computers. 
One approach to mitigate a related problem via distributed computing is described in TIGER~\cite{W2012} for large single genomes, though this approach has not been validated on metagenomic samples. The insight towards the composition of the metagenomic sample is also deferred until after assembly. One approach that mitigates this issue -- and the one pursued in this paper -- is to perform a rough classification of the reads into subgroups that may individually be processed more efficiently for the purpose of both identifying a proper reference and for subsequent compression. The structure of such an algorithm is depicted in Figure~\ref{fig:MCUIUC}, with a brief description of the blocks provided in what follows. 

\begin{itemize}
\item \textbf{Step 1 (Classification):} The first step of the procedure is a ``rough'' identification of the mixture of species in the metagenomic sample. One well established method for bacterial genome identification is the use of the so called 16S rRNA regions~\cite{Liu97}. Although this method is helpful for basic taxonomic identification\footnote{There are several taxonomy levels for describing living organisms, listed from most general to most specific: life, domain, kingdom, phylum, class, order, family, genus, and species.}, it does not allow for the precision of methods based on multiple marker identification, such as Metaphyler~\cite{Liuetal2010}.
The gist of the approach in~\cite{Liuetal2010} is that almost every genomic substring of length exceeding $20$ is unique to a species. Metaphyler hence scans for markers in excess of this length and links them to them to
markers of documented species. Therefore, known organisms have a high chance of being identified by this method up to a given taxonomic order, in this case, their genus. As will be shown in the next section, Metaphyler reports the number of identified markers of genera as well as their corresponding abundance levels, but the algorithm often misses a number of species (most often, undocumented species) while introducing a large number of false positives. In order to optimize the performance of the subsequent steps of the algorithm, the selection of genera most likely contained in the mixture has to be performed using the identified number of markers, the length of the genome and other criteria. We elaborate on this issue in the following section.
\item \textbf{Step 2 (Partitioning of the Dataset):} \begin{enumerate}
\item  Once a group of genera is selected according to the Procedure in Step 1, a set of reference genomes ``best'' aligned to the reads has to be selected. Given that Metaphyler produces predictions only up to the level of genera, one possible way to identify representative genomes in the mixture is to \emph{select all species of the identified genera} and perform quick testing of alignment quality with the given reads\footnote{We observe that a similar method using a small number of randomly selected genomes within a given genus does not produce good quality classification results.}. This task can be accomplished by using \emph{Bowtie2}~\cite{LS2012}, a specialized software for ultra-fast alignment of \emph{short} reads to \emph{long} genomes. Bowtie2 was designed as an alternative to the BLAST (Basic Local Alignment Search Tool)~\cite{A1990} for cases when one has a very large number of short query sequences to be aligned with a small number of genomes. Running Bowtie2 on the metagenomic reads and reference genomes from all species within the identified genera allows for performing read classification up to the genus level in terms of best alignment score. It is worth pointing out that some reads will be reported as unaligned, i.e., as not having sufficiently high similarity to any of the species to which alignment was performed. A means for parallelizing \emph{single genome} assembly that shares some of the classification ideas of Step 2 was first reported in~\cite{W2012}. 
\item In order to keep the number of reference genomes used for compression reasonably small, the next step is to combine the reads assigned to members within one single genus, independent on which species they were identified to belong to.  The species of each genus which had the most reads aligned to it is called the \emph{representative of the genus} and is used for reference-based compression of all the reads mapped to the genus. 
\item Unaligned reads are treated differently. Given that their number is relatively small compared to the number of aligned reads, metagenomic assembly of such reads may be performed in a computationally efficient manner using a metagenomic assembler such as IDBA-UD~\cite{Pengetal2012}. The assembler produces \emph{contigs} -- overlapped strings of reads of long lengths -- that may subsequently be queried in BLAST to identify the organisms they most likely originated from. Steps 2.1 and 2.2 are repeated using as input the unaligned reads from step 2.1 as well as the genera of organisms associated with the longest contigs.
\end{enumerate}
\item \textbf{Step 3 (Compression and Distribution):} The number of unaligned reads remaining after step 2.3 is relatively small and can be compressed using a standard tool such as bzip2. The reads associated with each genus are aligned to their representative using Bowtie2 and outputted into the SAM (Sequence Alignment/Map) format. The resulting SAM files are converted to the sorted and indexed BAM format (a binary format for sequencing data) via SAMtools~\cite{samtools}. The BAM file for each genus is compressed via reference-based compression against its representative to the CRAM format using the CRAM toolkit~\cite{cram}. The CRAM toolkit provides a practical reference-based compression procedure and file format based on~\cite{M2011}. Note a simple extension to the algorithm akin to~\cite{M2011} is to use some of the contigs as references for compression. These references would be packaged for distribution. 
The compressed unaligned reads, CRAM files, list of representatives for the genera, the corresponding genomes 
(if they are not available through a standard database such as NCBI's microbial genome database) and Metaphyler taxonomic classification are packaged into a tar archive. The resulting archive can be distributed and the reads losslessly reconstructed via the CRAM toolkit, given the CRAM files and representative genomes. The representative genomes may be compressed using standard compressors such as bzip2 or specialized compressors such as DNACompress~\cite{X2002}. 

\end{itemize}
\begin{table}[t!]
\caption{A randomly selected set of $15$ species used to illustrate the operating principles of MCUIUC.} 
\begin{center}
\begin{tabular}{||c|c|c|} \hline
$\#$ &	\textbf{Genus}	& \textbf{Species} \\ \hline
1 &	Aquifex	& Aquifex-aeolicus-VF5-uid57765 \\ \hline
2 &	Bacillus & 	Bacillus-amyloliquefaciens-FZB42-uid58271\\ \hline
3 &	Cycloclasticus &	Cycloclasticus-P1-uid176368 \\ \hline
4 &	Enterococcus	& Enterococcus-faecalis-D32-uid171261\\ \hline
5 &	Escherichia	& Escherichia-coli-S88-uid62979\\ \hline
6 &	Francisella	& Francisella-novicida-U112-uid58499\\ \hline
7 &	Geobacter	& Geobacter-FRC-32-uid58543\\ \hline
8 &	Helicobacter	& Helicobacter-pylori-P12-uid59327\\ \hline
9 &	Lactobacillus	& Lactobacillus-acidophilus-30SC-uid63605\\ \hline
10 &	Lactococcus	& Lactococcus-lactis-CV56-uid160253\\ \hline
11 &	Mycoplasma	& Mycoplasma-genitalium-G37-uid57707\\ \hline
12 &	Polynucleobacter	& Polynucleobacter-necessarius-STIR1-uid58967\\ \hline
13 &	Oscillatoria	& Oscillatoria\_PCC\_7112\_uid183110d\\ \hline
14 &	Thermococcus	&Thermococcus-4557-uid70841\\ \hline
15 &	Vibrio	& Vibrio-splendidus-LGP32-uid59353\\ \hline 
\end{tabular}
\end{center}
\label{default}
\vspace{-20pt}
\end{table}%

\vspace{-9pt}
\section{Working Example} \label{sec:example}
Metagenomic samples come in various sizes -- most environmental samples contain anywhere between tens to hundreds of species and subspecies. The human gut is estimated to contain roughly $300$ bacteria, while human reproductive systems host roughly $20-40$ microbes~\cite{Q2010}. In order to illustrate our findings in a compact form, we perform a step-by-step analysis of the compression scheme described in the previous section on simulated metagenomic data containing $15$ species.  

\subsection{Simulating the Metagenomic Sample}
Species were randomly selected from the NCBI microbial genome database available at~\cite{wp}. The selected group of $15$ organisms is listed in Table I.  Of the chosen species, Oscillatoria\_PCC\_7112\_uid183110 has the longest genome with $7,585,859$ bps, while Mycoplasma\_genitalium\_G37\_uid57707 has the shortest genome with $588,364$ bps.  For each species, we selected the FASTA file containing the complete genome and generate paired reads using the sim\_reads tool accompanying IDBA-UD with settings as in \cite{Pengetal2012}, and coverage depth $100$. The reads from each species were combined to simulate a metagenomic sample from an Illumina sequencer without the quality scores (though our algorithm may work with quality information as well, such as reads stored in the FASTQ format).  The resulting metagenomic sample was 5 GiB in the FASTA format. Note that real metagenomic data tend to have a more ``coherent'' set of species, i.e., a mix of many strains of the same organism and mixtures of organisms adapted to the same environmental conditions. This represents additional side information that may be exploited in the Identification stage of the procedure, but will not be elaborated on in this paper.

\subsection{Metaphyler Genus Identification}
Metaphyler takes the simulated metagenomic reads as inputs and outputs their taxonomy classifications. We focused our attention on genus classification, as it is the most specific level provided by Metaphyler.  The Metaphyler genus-level output for the input metagenomic reads is given in Fig. 4.  Of the $31$ genera identified, we selected only those with percent abundance higher than a threshold, which in this case was chosen as $0.1$. Note that the choice of the threshold is governed by many parameters, including the number of estimated organisms, their genome length, the number of known markers in the genomes, as well as the actual output of Metaphyler. A detailed analysis of the threshold selection method is relegated to the full version of the paper. The selected set of $14$ genera contains $13$ true positives and one false positive (note that Synechococcus was not part of the selected $15$ species).  In addition, Metaphyler missed identifying two organisms present in the metagenomic sample: Cycloclasticus\_P1\_uid176368 and Oscillatoria\_PCC\_7112\_uid183110.  

Given that Metaphyler produces only genus-level classifications, it remains to identify the species within the metagenomic mixture. We tried two approaches for addressing this problem: in the first method, we randomly selected a small number of species (2-8, or the largest number of documented species if this number was smaller than eight) for each of the identified genera. In the second approach, we selected \emph{all known species} within each chosen genera. As an illustration, more than $35$ strains of \emph{E. coli} are sequenced so far, with more than $20$ other sequencing efforts close to completion, while only one species of Cycloclasticus has been sequenced so far (see Wikipedia -- ecoliwiki -- for more information). In both cases, the genomes of the selected species were used as references for classification, as described in the next subsection.

\vspace{-5pt}
\subsection{Read Classification}
For the purpose of classifying the reads based on similarity to the selected reference genomes, we used the Bowtie2 algorithm~\cite{LS2012}. Bowtie2 is based on two concepts commonly encountered in computer science and information theory: the Burrows-Wheeler (BW) transform and suffix-tree searching. The BW transform is performed to index the reference genomes for subsequent read alignment. Searching for the best alignments is accomplished via suffix-trees, which makes the solution computationally efficient even for very large number of sequence reads. 

The classification results were of poor quality when only 2-8 reference genomes from each genera were selected randomly, and are not reported. On the other hand, using all species of the $14$ chosen genera above and building a Bowtie2 index provided very good metagenomic read alignment. In the first round of alignment, $78\%$ of reads were grouped according to the species they were identified to belong to. In the second round of alignment, only one species per genera with largest number of aligned reads was used as a reference for the genera. The unaligned reads from the first round were assembled using IDBA-UD. The resulting contigs were passed through BLAST. As an illustration, BLASTing the 30 longest contigs (of lengths 31k to 722k) led to identification of the two species missed by Metaphyler with very high confidence level: $12$ of the contigs came from Cycloclasticus\_NC018697, while $18$ contigs were identified as coming from Oscillatoria\_NC019729. 
Subsequently, unaligned read classification was performed for the two newly identified species, following the same steps as outlined in connection with the Metaphyler output processing scheme.

\begin{figure}[!t]
\centering
\includegraphics[width=3.0in]{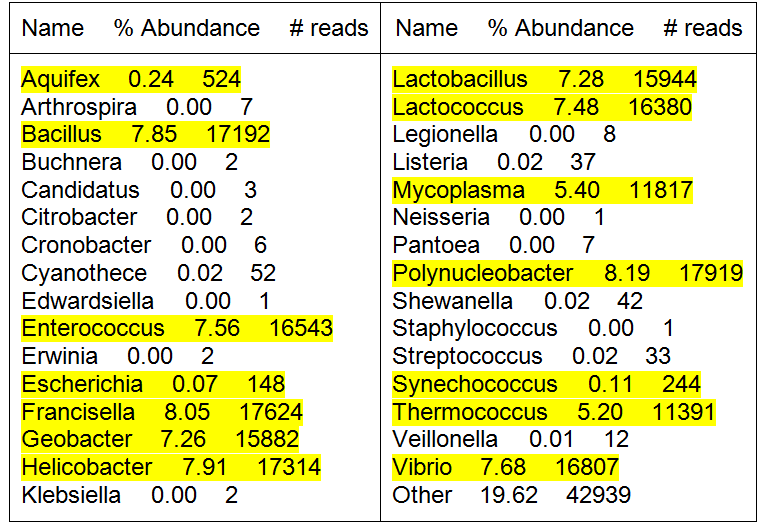}
\caption{Output of Metaphyler on randomly selected set of organisms, shown in Table 1. 14 genera with abundance higher than 0.1 are highlighted.}
\label{fig:metaphyler-output}
\end{figure} 
\subsection{Compression}
For comparison, the sample was initially compressed with gzip and bzip2 resulting in 1.4 GiB and 1.3 GiB file sizes, respectively. Both tools are common on UNIX-like platforms. Prior to CRAM conversion, the file stored in the sorted BAM format used 1.2 GiB while stored in the CRAM format used 0.4 GiB. The BAM format uses a variant of gzip internally~\cite{samtools}. Due to space constraints, details regarding the CRAM software are deferred to the specification~\cite{cram}. The compressed file is on the order of 10\% of the uncompressed file and 30\% of the file after bzip2 compression. Analogous to \cite{M2011}, we predict the gap will remain large as sequencing lengths increase and larger metagenomic samples are collected. 

As the number of reads mapped to Synechococcus is very small given the relative sizes of the detected genomes, it is reasonable to assume it was a false positive and lump the Synechococcus reads with the unaligned set for a clearer picture of the contents of the metagenomic sample. In addition, note that given the reference genome lengths, the proportion of each organism in the metagenomic sample can be estimated assuming uniform coverage. 

\begin{table}[htdp]
\caption{Size of reads mapped to each genera (Unit: nearest MiB)}
\begin{center}
\begin{tabular}{||l|r|r|r|}\hline
	&FASTA& BAM	&CRAM\\ \hline
Identified via Metaphyler & & &\\ \hline
Aquifex	&181	&45	&15\\
Bacillus	&458	&115	&40\\
Enterococcus	&349	&87	&30\\
Escherichia	&589	&149	&52\\
Francisella	&223	&56	&19\\
Geobacter	&504	&126	&44\\
Helicobacter	&195	&49	&17\\
Lactobacillus	&243	&60	&21\\
Lactococcus	&280	&70	&24\\
Mycoplasma	&67	&16	&6\\
Polynucleobacter	&182	&45	&16\\
Synechococcus	&1	&0	&0\\
Thermococcus	&235	&59	&20\\
Vibrio	&386	&96	&33\\ \hline
Identified via BLAST	&	&	&\\ \hline
Oscillatoria	&875	&218	&76\\
Cycloclastius	&276	&69	&24\\
Unaligned	 &1	&bzip2:	&0\\ \hline
Total	&5046	&1261	&436\\\hline
\end{tabular}
\end{center}
\label{compression}
\vspace{-20pt}
\end{table}

\section{Conclusion} \label{sec:conclusion}

We described the first \emph{de novo} metagenomic read compression algorithm. The algorithm was based on four major subroutines: genera identification, species identification, read classification and reference based single-genome compression. The performance of the method was illustrated on a simulated metagenomic sample with $15$ species, for which overall compression ratio on the order of 12.5 were reported. Further work includes integrating side information such as genera known to be present in a sample with high likelihood, selection rules for using multiple representatives for a genus and integration of phylogenic aligners to mitigate the ``Other'' category used in Metaphyler at the genus level. 

\section*{Acknowledgment}
This work was supported in part by NSF grants CCF 0809895, CCF 1218764, Emerging Frontiers for Science of Information Center, CCF 0939370 and U.S. Defense Threat Reduction Agency through subcontract 147755 at the University of Illinois from prime award HDTRA1-10-1-0086.
The authors also gratefully acknowledge many useful discussions with Prof. Jian Ma and Xiaolong Wu at the University of Illinois, Urbana-Champaign.

\end{document}